\documentclass[aps, prb, amsmath, amssymb, amsfonts, twocolumn, footinbib]{revtex4-1}
\usepackage[german,american,english]{babel}
\usepackage{graphicx}
\usepackage{graphics}
\usepackage{dcolumn}
\usepackage{bm}
\usepackage{amssymb}
\usepackage{amsmath}
\usepackage{amsfonts}
\usepackage{epsfig}
\usepackage{mathbbol}
\usepackage{latexsym}
\usepackage{dcolumn}
\usepackage{subfigure}
\usepackage{hyperref}
\usepackage{float}
\usepackage[normalem]{ulem}
\usepackage[usenames, dvipsnames]{color}
\usepackage{epstopdf}

\hypersetup{                    
    colorlinks,
    citecolor=blue,
    filecolor=blue,
    linkcolor=blue,
    urlcolor=blue
}

 \newcommand {\beq}{\begin{equation}}
\newcommand {\eeq}{\end{equation}}
\newcommand {\beqn}{\begin{eqnarray}}
\newcommand {\eeqn}{\end{eqnarray}}
\newcommand {\bit}{\begin{itemize}}
\newcommand {\eit}{\end{itemize}}

\newcommand{\ba}{\begin{array}{rl}}
\newcommand{\ea}{\end{array}}
\newcommand{\bc}{\begin{cases}}
\newcommand{\ec}{\end{cases}}

\newcommand{\om}{\iffalse}

\definecolor{mygray}{gray}{0.6}
\definecolor{gold}{RGB}{150, 150, 10}
\definecolor{mygreen}{RGB}{40, 200, 100}

\begin{document}
\title{Phase diagram of the interacting persistent spin-helix state}

\author{Hong Liu$^1$, Weizhe Edward Liu$^2$, Stefano Chesi$^3$, Robert Joynt$^4$ and Dimitrie Culcer$^1$}
\affiliation{$^1$School of Physics and Australian Research Council Centre of Excellence in Low-Energy Electronics Technologies, UNSW Node, The University of New South Wales, Sydney 2052, Australia}
\affiliation{$^2$School of Physics and Australian Research Council Centre of Excellence in Low-Energy Electronics Technologies, Monash Node, Monash University, Melbourne 3800, Australia}
\affiliation{$^3$Beijing Computational Science Research Center, Beijing 100193, China}
\affiliation{$^4$Physics Department, University of Wisconsin–Madison, Madison, Wisconsin 53706, USA}
\begin{abstract}
We study the phase diagram of the interacting two-dimensional electron gas (2DEG) with equal Rashba and Dresselhaus spin-orbit coupling, which for weak coupling gives rise to the well-known persistent spin-helix phase. We construct the full Hartree-Fock phase diagram using a classical Monte-Carlo method analogous to that used in Phys.Rev.B \textbf{96}, 235425 (2017). For the 2DEG with only Rashba spin-orbit coupling it was found that at intermediate values of the Wigner-Seitz radius $r_s$ the system is characterized by a single Fermi surface with an out-of-plane spin polarization, while at slightly larger values of $r_s$ it undergoes a transition to a state with a shifted Fermi surface and an in-plane spin polarization. The various phase transitions are first-order, and this shows up in discontinuities in the conductivity, and the appearance of anisotropic resistance in the in-plane polarized phase. In this work we show that the out-of-plane spin polarized region shrinks as the strength of the Dresselhaus spin-orbit interaction increases, and entirely vanishes when the Rashba and Dresselhaus spin-orbit coupling strengths are equal. At this point the system can be mapped onto a 2DEG without spin-orbit coupling, and this transformation reveals the existence of an in-plane spin polarized phase with a single, displaced Fermi surface beyond $r_s > 2.01$. This is confirmed by classical Monte-Carlo simulations. We discuss experimental observation and useful applications of the novel phase, as well as caveats of using the classical Monte-Carlo method.
\end{abstract}
\date{\today}
\maketitle

\section{Introduction}\label{Intro}
The two-dimensional electron gas (2DEG) with spin-orbit coupling and many-body electron-electron interactions is a paradigmatic system in semiconductor physics and technology, in addition to being one of the fundamental models in condensed-matter physics. Much of the interest in spin-orbit coupling centers around the fact that it enables spin generation, spin manipulation and spin detection without using external magnetic fields or magnetic materials \cite{Ghosh04prl,Bihlmayer15newjphys}, while at the same time being manifest in a great variety of spin textures in solids \cite{Hsieh-Science-2008, Attaccalite-PRL-2002, RevModPhys.89.011001}, many of which are associated with topological effects \cite{Jiaji_Zhu_PRL_2011,Shekhter05prb,PhysRevB.74.115126}, unconventional states of matter \cite{Fu15prl, Berg2012} and novel phases \cite{Tanatar-PRB-1989, Baguet-PRL-2013, Ghosh-PRL-2004, Juri-PRB-2008, PhysRevB.84.115115, Drummond2009, PhysRevB.67.073304, PhysRevB.15.2819, PhysRevB.18.3126, PhysRevLett.82.5317, PhysRevLett.90.136601}. Acquiring a full understanding of the spin-orbit coupled 2DEG is key to our ability to utilise the electron spin degree of freedom in semiconductors to control the spin states and transfer spin information, which is a fundamental requirement for future spintronic devices and quantum computing \cite{Datta90apl, Zutic04rmp, Sinova_SHE_RMP15, Shen14prl, Jiaji-PRB-2010} among other applications. 

Keeping in mind both basic science and potential technological interest, identifying many-body ground states with novel spin textures and polarizations is one of the goals of present-day condensed matter research. This problem is notoriously difficult analytically even in the absence of spin-orbit coupling. The Hartree-Fock (HF) method often provides simple analytical solutions, and though it entirely ignores the effect of correlations it generally provides useful insights into the structure of the single-particle levels \cite{PhysRevB.83.235309, PhysRevB.60.4826, PhysRevB.72.035114}. In addition to this, the past few decades have seen dramatic improvements in our ability to simulate complicated physical systems using Monte Carlo (MC) approaches \cite{Foulkes-RMP-2001,Braguta-PRB-2016,Ceperley-PRL-1980,Ambrosetti-PRB-2009,Sam-2017,Kapfer-PRE-2016,Kensaku-2016}. At the same time, the transition of the electron gas into either a partially or a fully polarized fluid is not fully understood \cite{Tanaka-PRB-1989,Tanatar-PRB-1989,Attaccalite-PRL-2002,Varsano_2001}. At variance with the 3D case, there is the possibility of a partially polarized phase \cite{Tanaka-PRB-1989}, while for the 2D case there is no evidence for a stable partially polarized phase \cite{Tanatar-PRB-1989}. Furthermore, calculations at intermediate densities are difficult because very small energy differences are important and any approximation has to treat the various phases of the gas with equal accuracy \cite{Ceperley_nature}. 

The situation becomes even more complicated in spin-orbit coupled 2DEGs. In general the spin-orbit coupling (SOC) in semiconductor 2DEGs can take several forms. In realistic semiconductor nanostructures both the Rashba and the Dresselhaus spin-orbit coupling are often present. Rashba SOC is present primarily because quantum wells frequently have a built-in asymmetry \cite{Rashba}, and has been experimentally observed in semiconductor heterostructures, where it has been proved to be tunable in strength by means of a gate voltage \cite{TuneRashba3,TuneRashba2,TuneRashba1}. The Dresselhaus SOC reflects the inversion asymmetry inherent in zincblende lattices, which includes the crystal structure of many III-V and II-VI semiconductors such as GaAs, InSb, and CdTe \cite{Dresselhaus}. In 2DEGs the SOC can be described by an effective momentum-dependent magnetic field.  This effective field favors spin textures that have zero net moment, while the electron-electron interactions favor ferromagnetism.  This gives rise to a multitude of phases, many of which remain to be explored in detail. An insightful theoretical approach adopted in earlier studies of interacting spin-orbit coupled systems involved applying a unitary transformation to leading order in the spin-orbit strength, which yields a transformed Hamiltonian whose eigenstates are also spin and angular momentum eigenstates \cite{Aleiner-PRL-2001, PRB-2002-U}. More recently it has been shown that in addition to the well known out-of-plane spin polarized phase, \cite{Chesi_arXiv, Juri-PRB-2008} the interacting 2DEG with Rashba spin-orbit coupling exhibits an in-plane spin-polarized phase with a shifted Fermi surface \cite{Weizhe_Phase_PRB_2017}, somewhat resembling a Pomeranchuk instability.  It is caused by an exchange enhancement of the current-induced spin polarization. This phase appears already at intermediate values of $r_s$, the Wigner-Seitz radius, which represents the relative strength of the electron-electron interactions to the average kinetic energy. The same result is expected in systems with Dresselhaus SOC, since the Rashba and Dresselhaus interactions are related by a spin rotation. Several additional recent works have examined the interplay between Rashba spin-orbit coupling and electron-electron interactions in 2DEGs \cite{Baguet-PRL-2013, PhysRevB.84.115115, Marinescu_PRB15}.

Motivated by these observations we examine the phase diagram of the interacting 2DEG with Rashba and Dresselhaus spin-orbit interactions, focusing on systems with nearly equal Rashba and Dresselhaus SOC \cite{Walser-Nat-Phys-2012,Koralek-Nat-2009,Dettwiler-PRX-2017,Bernevig-ShouCheng-PSH-2006}.  The system with exactly equal Rashba and Dresselhaus interactions is an interesting special case with SU(2) symmetry. This symmetry is robust against spin-independent disorder and interactions, and is generated by operators whose wave vector depends on the coupling strength. It renders the spin lifetime infinite at this wave vector, giving rise to a persistent spin helix, \cite{Bernevig-ShouCheng-PSH-2006,  PhysRevB.86.081306, Koralek-Nat-2009} which has been realised experimentally \cite{Koralek-Nat-2009, Walser-Nat-Phys-2012, Dettwiler-PRX-2017}. When the Rashba and Dresselhaus interactions are of equal magnitude, the effective magnetic field describing the spin-orbit interaction singles out a well-defined direction in momentum space. There is a single spin-quantization axis, and all the spins in the system point either parallel or antiparallel to this axis. We determine the full Hartree-Fock phase diagram of the 2DEG with equal Rashba and Dresselhaus interactions. 

When either the Rashba or the Dresselhaus interaction is dominant we expect the same phase diagram as in Ref.~\onlinecite{Weizhe_Phase_PRB_2017}, while the case when the two interactions are equal is qualitatively different. Experimentally, the phase transitions are observe in transport properties, most notably the DC conductivity.  We present analytical results for this case and perform classical MC simulations along the lines of Ref.~[\onlinecite{Weizhe_Phase_PRB_2017}], focusing on the zero-temperature properties of the electron gas. Our analytical calculations reveal that the out-of-plane spin polarized phase shrinks as the magnitudes of the Rashba and Dresselhaus interactions approach each other, \textit{i.e.}, as the persistent spin helix state is approached. When the Rashba and Dresselhaus spin-orbit couplings are equal the out-of-plane phase disappears altogether and only an in-plane spin-polarized phase exists for all $r_s>2.01$. It is characterized by a single Fermi surface, which is displaced from the Brillouin zone center, and has a non-trivial spin texture, which becomes more pronounced at higher values of $r_s$ and of the spin-orbit interaction strength. Our expectations based on this analytical treatment are confirmed by classical MC simulations, which rely on the same method as that used in Ref.~\onlinecite{Weizhe_Phase_PRB_2017}. 

We find, however, two caveats related to the application of the classical MC method to this system. Firstly, in a narrow parameter regime the \textit{bare} MC phase diagram appears to display a phase with a small out-of-plane spin polarization, which we have referred to as OP$^*$ throughout this manuscript. To determine whether this phase is physical or not we varied the number of points in the MC simulations and studied the evolution of the ground state in this region as a function of the number of k-points $N$. We have found that in the thermodynamic limit $N \rightarrow \infty$ the ground state has an in-plane spin polarization in exact agreement with the analytical calculations. The energy difference, while larger than the numerical error, remains relatively small. Secondly, the persistent spin helix system exhibits a spin-density wave phase that is degenerate with the in-plane spin-polarized phase and that is not captured accurately by our MC method. The existence of this phase can be determined by analytical arguments, and we believe it is related to the appearance of the OP$^*$ phase in the MC phase diagram.  The application of the classical MC depends on using a class of HF wavefunctions for which the ground state energy can be reduced to a classical minimization problem.  The method will not find ground states outside this class such as the spin-density wave.  
 \begin{figure}
\begin{center}
\includegraphics[width=1\columnwidth]{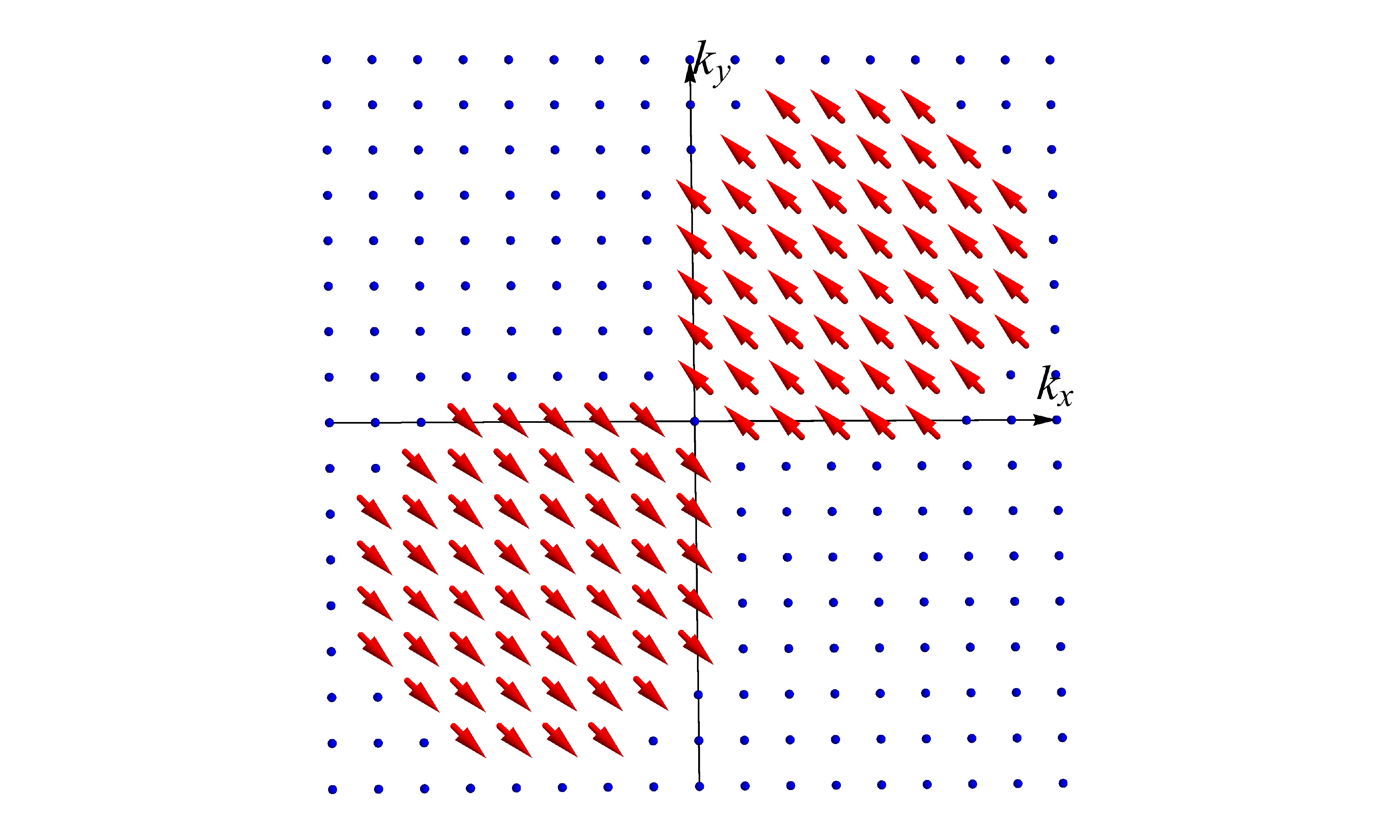}
\caption{\label{2FSs} Two-Fermi surface state at $\tilde{\alpha}=\tilde{\beta}=0.81$, $r_s=1.65$.}
\end{center}
\end{figure}
 \begin{figure}
\begin{center}
\includegraphics[width=1\columnwidth]{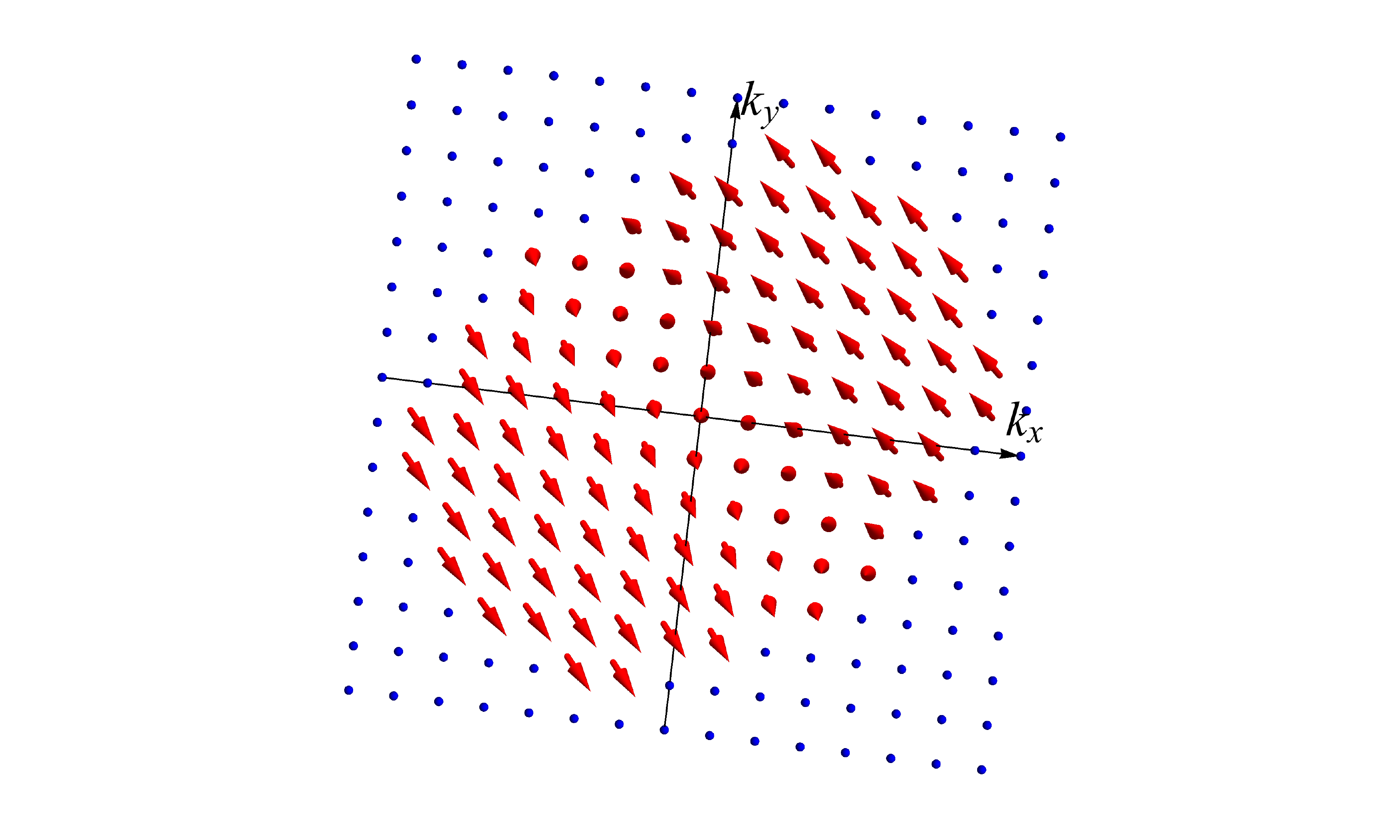}
\caption{\label{OP-0P45} OP$^*$ state at $\tilde{\alpha}=\tilde{\beta}=0.45$, $r_s=2.10$.}
\end{center}
\end{figure}

The organization of this paper is as follows: In Sec.~\ref{Analysis}, we describe the Hamiltonian of the system and the spin texture as well as the phase diagram obtained from analytical arguments using a gauge transformation. In Sec.~\ref{Ground-states}, the phase diagram of the interacting 2DEG with equal Rashba and Dresselhaus spin-orbit coupling is given by our classical MC simulation. In Sec.~\ref{Experimental Verification}, we compare the results obtained from the two preceding sections and discuss the possibilities for experimental observation. In Sec.~\ref{Conclusion}, we summarize our results.

\section{Hamiltonian and gauge transformation}\label{Analysis}

The general many-body Hamiltonian including both Rashba and Dresselhaus SOC reads: 
\begin{equation}\label{N-H}
H_{\alpha, \beta}=\sum_{{\bf k}ss'}\langle {\bf k} s|H^{(0)}_{\alpha, \beta}| {\bf k} s'\rangle c^{\dag}_{{\bf k},s}c_{{\bf k},s'}+V^{ee},
\end{equation}
where $c_{{\bf k},s}$ is the annihilation operator for a single electron with wave vector ${\bf k}$ and spin index $s =\pm$, and $c^{\dag}_{{\bf k},s}$ is the corresponding creation operator. Here we consider a translationally invariant system,  which is permissible as long as localization effects are negligible. Hence, only diagonal matrix elements in ${\bf k}$ appear in the single-particle Hamiltonian (first term). Including both Rashba ($\alpha$) and Dresselhaus ($\beta$) spin-orbit couplings, $H^{(0)}_{\alpha, \beta}$ reads:
\begin{equation}\label{Band-H}
H^{(0)}_{\alpha, \beta}= \frac{p^2}{2m}+\alpha(p_y\sigma_x-p_x\sigma_y)+\beta (p_x\sigma_x-p_y\sigma_y),
\end{equation}
where  $\sigma_i$ are Pauli  matrices. The electron-electron interaction $V^{ee}$ takes the standard form
\begin{equation} \label{Vee}
V^{ee}=\frac{1}{2A}\sum_{{\bf k}{\bf k}'ss'}\sum_{{\bf q}\neq0}V_{\bf q}c^{\dag}_{{\bf k}+{\bf q},s}c^{\dag}_{{\bf k}'-{\bf q},s'}c_{{\bf k}'s'}c_{{\bf k}s}.
\end{equation}
where $A$ is the area of the system. The specific form of $V_{\bf q}$ is not important for the the arguments developed in this section, as long as $V^{ee}$ has the usual coordinate representation $V^{ee} = \sum_{i<j} V({\bf r}_i - {\bf r}_j)$ [which is the case for Eq.~(\ref{Vee})]. Later we will use the screened 2D Coulomb interaction, which gives the following matrix element for momentum transfer ${\bf q}$:
\begin{equation}
V_{\bf q}=\frac{e^2}{2\varepsilon_r\varepsilon_0(k_{\text{TF}}+|{\bf q}|)},
\end{equation}
where $k_{\text{TF}}$ is the Thomas-Fermi wave number and $\varepsilon_r$ is the static dielectric constant. 

We now specialize Eq.~(\ref{N-H}) to the main case of interest in this paper, with equal Rashba and Dresselhaus coefficients ($\alpha = \beta$). Then it is convenient to rewrite the single-particle Hamiltonian Eq.~(\ref{Band-H}) as
\begin{equation}\label{H0_tilde}
H^{(0)}_{\alpha}=\frac{\tilde{p}^2_x+\tilde{p}^2_y}{2m}- 2\alpha \tilde{p}_x \tilde{\sigma}_y,
\end{equation}
where $\tilde{p}_x=(p_x+p_y)/\sqrt{2}$ and $\tilde{p}_y=(p_y-p_x)/\sqrt{2}$ are rotated by $\pi/4$ with respect to the original coordinates, and the spin operators $\tilde{\sigma}_{x,y}$ defined in a similar manner ($\tilde{\sigma}_z=\sigma_z$). Importantly, the natural spin quantization axis is along a fixed direction
$ {\bf e}_{\tilde{y}} = ({\bf e}_{y}-{\bf e}_{x})/\sqrt{2}$, independent of $\bf k$. The single-particle spectrum is immediately obtained as
\begin{equation}\label{eps_SOI}
 \varepsilon_{0,{\bm k}\pm}=\frac{\hbar^2 ({\bm k}\mp{\bm q}_\alpha)^2}{2m}-2m\alpha^2,
\end{equation}
where ${\bf q}_\alpha= \frac{\sqrt{2}\alpha m}{\hbar} ({\bf e}_x+{\bf e}_y)$. Note that the $\pm {\bf q}_\alpha$ shift in momentum appearing in Eq.~(\ref{eps_SOI}) gives rise to two distinct Fermi surfaces, displaced in opposite directions.

Besides the existence of the conserved quantity $\tilde{\sigma}_y$, it was recognized early on that a spatially-dependent spin rotation:
\begin{equation}
    U_\alpha = \exp\left[ i\frac{m \alpha}{\hbar} (\sigma_y - \sigma_x)(x+y)\right],
\end{equation}
relates the non-interacting Hamiltonain $H^{(0)}_{\alpha}$ to the familiar case without SOC, and this is still true after including spin-independent potentials. \cite{Schliemann2003, Bernevig-ShouCheng-PSH-2006} The many-body form of $U_\alpha$ acts as $U_\alpha  c_{{\bf k},s} U_\alpha^\dag =  c_{{\bf k}+s {\bf q}_\alpha,s} $ and commutes with the electron-electron interaction. Therefore, the whole family of many-body Hamiltonians $H_\alpha = H_{\alpha,\alpha}$ may be related to the extensively studied 2D electron liquid without SOC:\cite{Vignale.05}
\begin{equation}\label{halpha}
H_\alpha= U_\alpha \left( H_{\alpha= 0} \right) U_\alpha^\dag - 2 N_e m \alpha^2,
\end{equation}
where  $N_e$ is the total number of electrons.

Many properties of the spin-orbit coupled system can be obtained directly from the exact mapping (\ref{halpha}), and here we will be interested in the occurrence of a spontaneous spin polarization. It is then useful to consider the spin density operators $\tilde{{\bf S}}({\bf r})= \sum_i \tilde{{\boldsymbol{\sigma}}}_i \delta({\bf r}-{\bf r}_i) $, where $i=1,\ldots N_e$ labels the electrons and $\tilde{\boldsymbol{\sigma}}$ are the rotated Pauli matrices introduced after Eq.~(\ref{H0_tilde}). $U_\alpha$ transforms the spin polarization as follows:
\begin{align}
    U_\alpha \tilde{S}_x({\bf r})  U_\alpha^\dag &= \tilde{S}_x ({\bf r})\cos 2 {\bf q}_\alpha {\bf r} +  S_z({\bf r}) \sin 2 {\bf q}_\alpha {\bf r}, \label{Sx_transf}\\
    U_\alpha S_z({\bf r})  U_\alpha^\dag &= S_z ({\bf r})\cos 2 {\bf q}_\alpha {\bf r} - \tilde{S}_x({\bf r}) \sin 2 {\bf q}_\alpha {\bf r},\label{Sz_transf}
\end{align}
while  $U_\alpha \tilde{S}_y({\bf r})  U_\alpha^\dag = \tilde{S}_y({\bf r}) $. Therefore, a uniformly polarized state without spin-orbit interaction leads to a spin wave dependence when $\alpha \neq 0$. The only exception is a polarization along ${\bf e}_{\tilde{y}}$, which is kept unchanged. 

A family of degenerate spin waves is obtained from the SU(2) symmetry. In the system without spin-orbit interaction, the symmetry operations are simply spin rotations. However, combining an arbitrary spin rotation of the state uniformly polarized along ${\bf e}_{\tilde y}$ with the $U_\alpha$ transformation will induces a spatial precession of the $\tilde{S}_{x,z}$ components. These collective spin-waves can become the relevant ground states through the combined action of electron-electron interactions and spin-orbit coupling. In the same way that a Stoner transition leads to ferromagnetism (with $\alpha =0$), a sufficiently strong electron interaction will lead to the spontaneous formation of the spin-wave states  (when $\alpha \neq 0$). In this sense, the Stoner transition without spin-orbit interaction corresponds to the spontaneous formation of collective persistent spin-helix states.

The paramagnetic phase is apparently less interesting, as there is no spin polarization. However, the existence of persistent spin-helix states is reflected on the quasiparticle excitations. We first note that, without spin-orbit coupling, the spin-degenerate Fermi surface survives the effect of the interactions (the Luttinger's theorem \cite{Vignale.05}). Applying $U_\alpha$ to the interacting state without spin-orbit coupling leads to two distinct Fermi surfaces, centered around $\pm {\bf q}_\alpha$. Like for the non-interacting case [see  Eq.~(\ref{eps_SOI})], the two Fermi surfaces correspond to orthogonal spin directions $\pm$ along $\tilde \sigma_y$. 

If now we consider quasiparticle excitations, at $\alpha=0$ they can be generated by single-particle operators with arbitrary spin direction,  $\cos{(\theta/2)}c^{\dag}_{{\bf k},+} +e^{i\phi} \sin{(\theta/2)}c^{\dag}_{{\bf k},-}$ where $k\simeq k_{F}$ is close to the Fermi wavevector  ($k_F=\sqrt{2\pi n}$, where $n$ is electron density). For $\alpha \neq 0$, these operators transform to the following form: 
\begin{equation}\label{ck_soi}
c^{\dag}_{{\bf k}, \hat{n}} \equiv\cos{(\theta/2)}c^{\dag}_{{\bf k}+{\bf q}_\alpha,+} +e^{i\phi} \sin{(\theta/2)}c^{\dag}_{{\bf k}-{\bf q}_\alpha,-},
\end{equation}
which allows us to define stable quasiparticles $c^{\dag}_{{\bf k},{\hat{n}}} |F\rangle$ on top of the (interacting) ground state $| F\rangle$. 

From the previous discussions, especially Eqs.~(\ref{Sx_transf}) and (\ref{Sz_transf}), it is clear that the quasiparticles states $c^{\dag}_{{\bf k},{\hat{n}}} |F\rangle$ carry a spatially precessing spin polarization. In fact, the form of Eq.~(\ref{ck_soi}) involves a coherent superpositions of states at both Fermi surfaces (except for $\theta=0$), and the $\pm {\bf q}_\alpha$ displacements induce the formation of a spin-wave with wavevector $2{\bf q}_\alpha$. For a Fermi liquid, the lifetime of the $c^{\dag}_{{\bf k},{\hat{n}}} |F\rangle$ states becomes infinite in the limit $k\to k_F$. Then, we see that in the paramagnetc phase the natural excitations of the system are interacting persistent spin helix states.

With the help of the $c^{\dag}_{{\bf k}, \hat{n}}$ operators, we can give a mean-field description of the Stoner transition with spin-orbit interaction. The paramagnetic phase corresponds to:
\begin{equation}
|F \rangle \simeq \prod_{k\leq k_F} c^{\dag}_{{\bf k}, \hat{n}}c^{\dag}_{{\bf k}, -\hat{n}} |0\rangle,
\end{equation}
where the choice $\hat n = {\bf e}_{\tilde y}$ is perhaps more natural, but any other direction of $\hat n $ is equivalent: they all give the same spin-unpolarized ground state of the non-interacting Hamiltonian. On the other hand, the mean-field spin-polarized states can be written as:
\begin{equation}\label{GS_polarized}
|F \rangle \simeq \prod_{k \leq \sqrt{2} k_F} c^{\dag}_{{\bf k}, \hat{n}}  |0\rangle,
\end{equation}
which actually depend on the direction $ \hat{n}$, reflecting the broken SU(2) symmetry.  As discussed, only when $\hat n = {\bf e}_{\tilde y}$ the spin density is uniform. For other orientations, all the electrons occupy spin-helix states and the spin density is precessing in space as described by Eqs.~(\ref{Sx_transf}) and (\ref{Sz_transf}). It is oriented along $\hat{n}$ only at periodic positions along the ${\bf e}_{\tilde x}$ direction (e.g., $\hat{n}$ can be taken as the polarization direction at ${\bf r}=0$).
 
For the unscreened Coulomb interaction, the phase diagram of the electron liquid at a fixed ratio $\beta/\alpha$  depends on two dimensionless parameters. The Wigner-Seitz radius $r_s$ is
\begin{equation}
r_s=\frac{me^2}{4\pi \varepsilon_r\varepsilon_0\hbar^2\sqrt{\pi n}},
\end{equation}
and is a measure of the interaction strength.\cite{Vignale.05} Instead, the Rashba spin-orbit coupling can be rescaled as follows:\cite{Weizhe_Phase_PRB_2017}
\begin{equation}
\tilde{\alpha}=\frac{m\alpha}{\hbar\sqrt{\pi n}}.
\end{equation}
Here, $\tilde{\alpha}$ is a measure of the Rashba spin-orbit coupling relative to the kinetic energy. While for $\beta=0$ the dependence of the phase boundaries on $\tilde{\alpha}$ is non-trivial,\cite{Weizhe_Phase_PRB_2017} when $\alpha=\beta$ we find from Eq.~(\ref{halpha}) that the phase boundares should be independent of $\tilde{\alpha}$. They are simply given by the values in the absence of spin-orbit interaction. 

\begin{figure}
\begin{center}
\includegraphics[width=8cm]{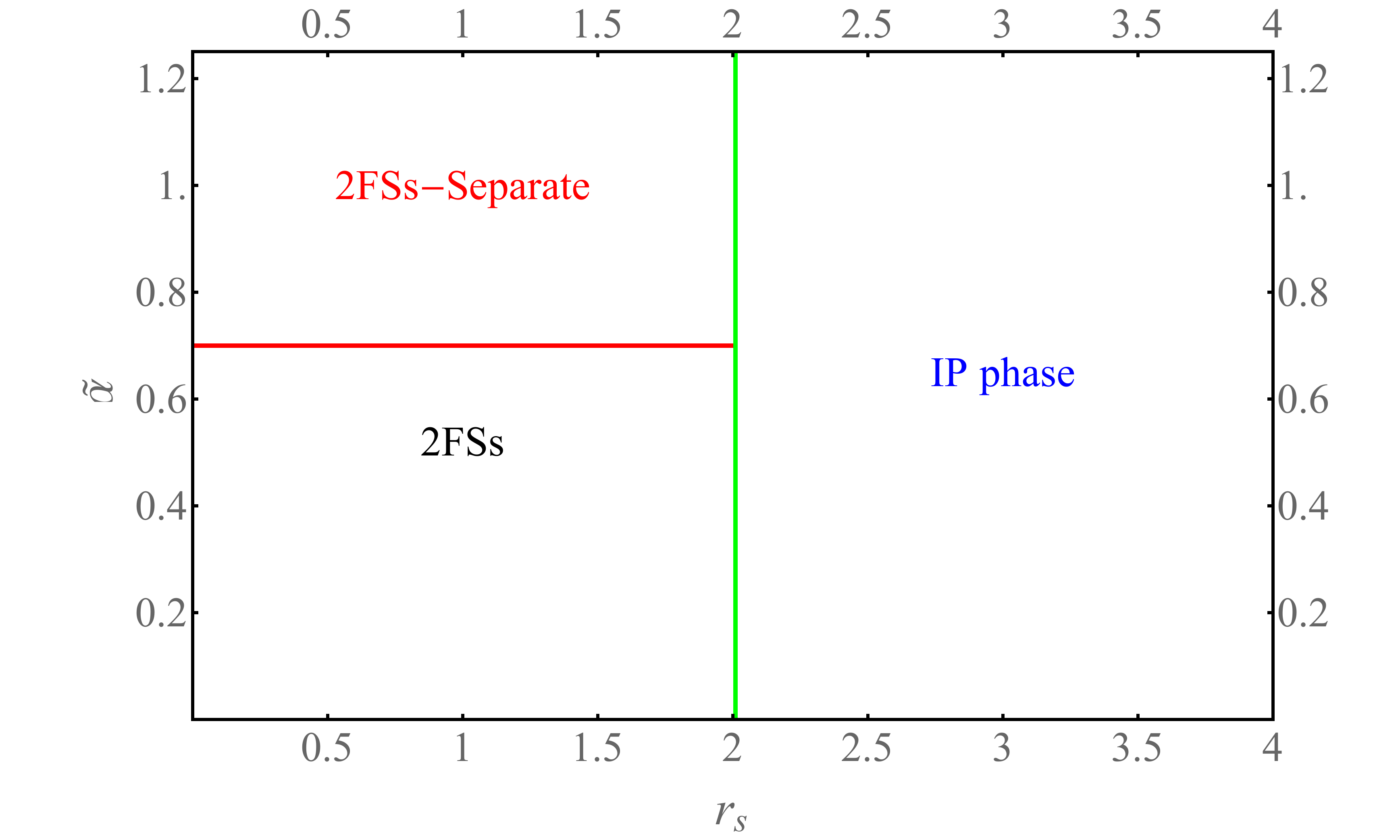}
\caption{\label{Boundary} Phase diagram of 2D electron liquid with Rashba equal to Dresselhaus spin-orbit coupling in the Hartree-Fock approximation, obtained by gauge transformation. Here, $\tilde{\alpha}$ and $r_s$ are dimensionless measures for the strength of Rashba (Dresselhaus) spin-orbit coupling and electron-electron interactions, respectively. For the paramagnetic Fermi-liquid phase 2FSs and 2FSs-separate, there is no net spin polarization. The IP phase exhibits an in-plane magnetization associated with a shifted Fermi sea.}\label{fig1}
\end{center}
\end{figure}

For example, it is well known that the Hartree-Fock approximation gives a Bloch transition at $r_s \simeq 2.01$ and we show in Fig.~\ref{fig1} the simple mean-field phase diagram, where the transition to the states of type of Eq.~(\ref{GS_polarized}) is independent of $\tilde\alpha$. In the paramagnetic region, $r_s < 2.01$, we also mark the region  $\tilde\alpha > 1/\sqrt{2}$ in which the $s=\pm$ Fermi discs become non-overlapping (obtained from the condition $|{\bf q}_\alpha| > k_F$).

Finally, we comment on the average value of the momentum operator ${\bf P} = \sum_i {\bf p}_i$ in the collective persistent spin helix states, giving:
\begin{equation}\label{P_average}
   \frac{1}{N_e} \langle F |{\bf P}| F \rangle =   \hbar{\bf q}_\alpha \cos\theta.
\end{equation}
Except for $\theta=\pi/2$, there is a finite expectation value of ${\bf P}/N_e$ along ${\bf e}_{\tilde x}=({\bf e}_x+{\bf e}_{y})/\sqrt{2}$, reflecting a displaced momentum-space occupation of the spin-polarized states. The displacement is largest when the polarization is uniform and oriented along $\tilde{\sigma}_y$ ($\theta=0$). As usual, Eq.~(\ref{P_average}) does not imply a finite current in equilibrium, since the single particle velocity along ${\bf e}_{\tilde x}$ is given by $\tilde{p}_x/m -2 \alpha \tilde{\sigma}_y$. The finite expectation value of ${\bf P}/N_e$ is exactly cancelled by the contribution from the spin polarization:
\begin{equation}
    \frac{2 m \alpha}{n} \langle F |\tilde{S}_y({\bf r})| F \rangle =  2m\alpha \cos\theta .
\end{equation}
Note from Eqs.~(\ref{Sx_transf}) and (\ref{Sz_transf}) that $\langle F |\tilde{S}_y({\bf r})| F \rangle = n \cos{\theta}$ is the only component of the spin polarization density which is independent on ${\bf r}$.

\section{Monte Carlo simulation of interacting 2DEG with Rashba and Dresselhaus spin-orbit coupling}\label{Ground-states}
In the statically screened Hartree-Fock approximation, the exchange energy of the system can be written as
\begin{equation} 
E_\text{ex} =-\frac{1}{L^2}\sum_{{\bm k}\neq {\bm k}'} \frac{e^2[s_{\bm k}\cdot s_{{\bm k}'}+n_{\bm k} n_{{\bm k}'}]}{4\epsilon_0 (k_{\text{TF}}+|{\bm k}-{\bm k}'|)},
 \end{equation}
  where $f_{\bm k}$ is the density matrix in equilibrium, while
 $n_{\bm k}=\frac{1}{2}\text{Tr} f_{\bm k}$ and  $s_{\bm k}=\frac{1}{2}\text{Tr} ({\bm \sigma}f_{\bm k})$ are the electron's occupation number and net spin polarization at ${\bm k}$, respectively. The total energy can be expressed as
  \begin{equation}
E_{\text{tot}}=\text{Tr}[f_{\bm k}H_{0{\bm k}}]+E_\text{ex},
\end{equation}
 We assume that the screening effect is negligible due to the low electron density, so $k_\text{TF}=0$. In the following subsections we use the same method as in  Ref.~[\onlinecite{Weizhe_Phase_PRB_2017}] to find the minimum-energy configuration. When $r_s$ is close to a phase boundary, the total energy is linearly dependent on $r_s$, which allows us to use a linear fitting to determine the transition points.

\subsection{Simulation of 2DEG with Rashba spin-orbit coupling}\label{2DEG+R}
Using a MC simulation, we reproduce the result for 2DEG with Rashba spin-orbit coupling in Ref.~[\onlinecite{Weizhe_Phase_PRB_2017}]. As $r_s$ increases, the interaction becomes more effective, producing a tendency towards ferromagnetism. When $\tilde{\alpha}=0$ there is the classic Bloch transition that occurs at $r_s=2.01$. 
There are two conventional Fermi liquid (FL) states with one and two occupied spin sub-bands, respectively. The only effect of the exchange interaction is to renormalize upwards the strength of the Rashba term, and there is no net spin polarization for the two Fermi liquid states. The phase boundary of the two Fermi liquid states is well described by the (noninteracting) critical density equation, $n_c=\frac{m^2\alpha^2}{\pi \hbar^4}$. As $r_s$ increases with finite $\tilde{\alpha}$,  the ferromagnetic phase is modified to the OP phase, where spins have a $z$ component and a component along the effective field due to the Rashba spin-orbit coupling. At small $k$, they point nearly along the $z$ direction, but as $k$ increases, they follow the spin orbit-induced field. When $r_s$ is even larger,  the right half of the phase diagram is the IP phase.  The key feature of the IP phase is that the spin polarization is completely in-plane and the IP phase does not have any symmetry on the Fermi surface, even though both of them only have a single band.

\subsection{Simulation of 2DEG with Rashba and Dresselhaus spin-orbit coupling}
\begin{figure}
\begin{center}
\includegraphics[width=9cm]{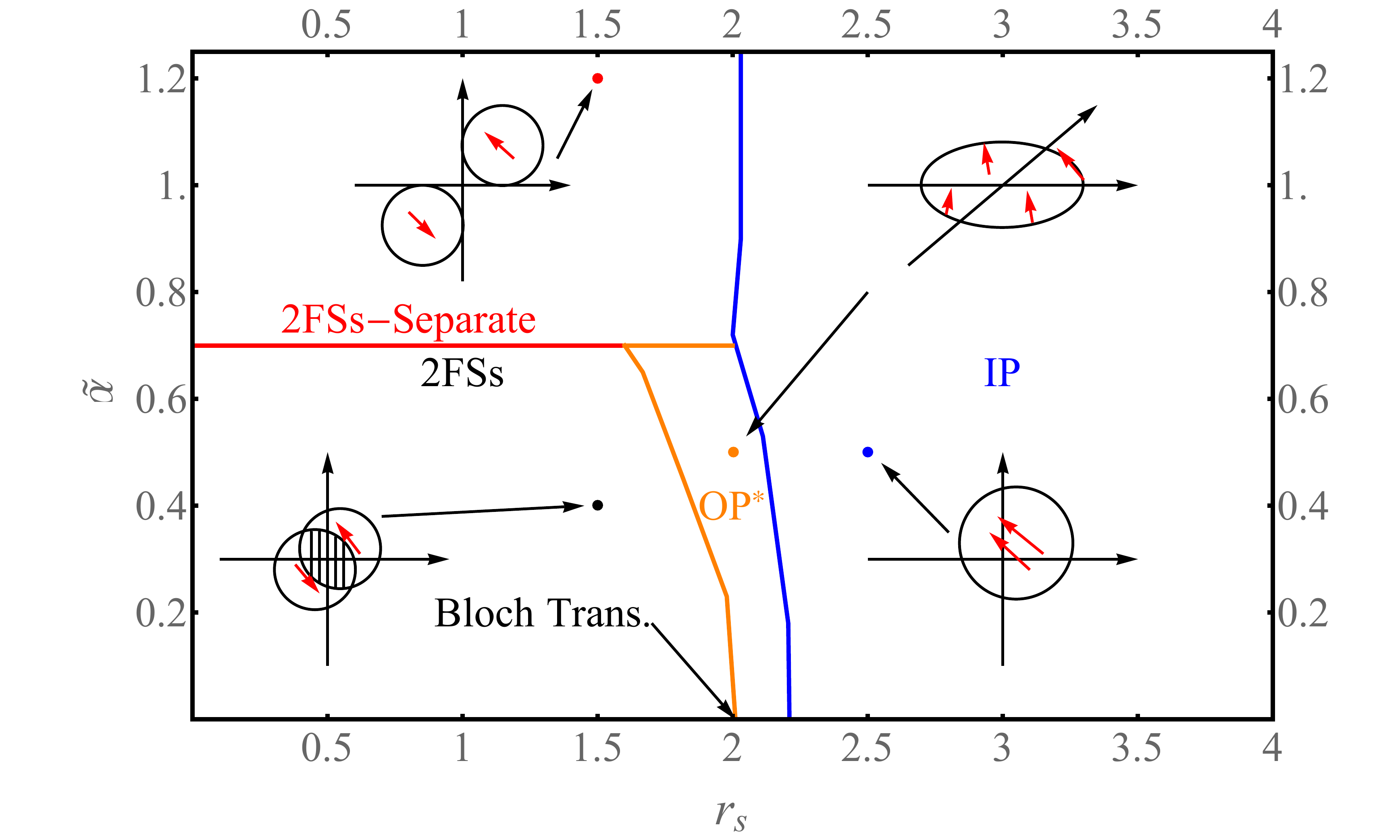}
\caption{\label{B-MCS} Phase diagram of a 2D electron liquid with equal Rashba and Dresselhaus spin-orbit couplings, obtained by solving the Hartree-Fock equations using a Monte Carlo method. Here, $\tilde{\alpha}$ and $r_s$ are dimensionless measures for the strength of Rashba spin-orbit coupling and the electron-electron interactions, respectively: 
$\tilde{\alpha}$  corresponds to the ratio of the Fermi wavelength and spin-precession length, and $r_s$ is the Wigner-Seitz radius of the 2D electron system. The distinguishing features for each individual phase are indicated schematically. For the Fermi-liquid phases 2FSs and 2FSs-separate, there is no net spin polarization. In contrast, the OP$^*$ phase is characterized by an out-of-plane magnetization. The IP phase exhibits an in-plane magnetization associated with a shifted Fermi sea.}
\end{center}
\end{figure}

\begin{figure}
\begin{center}
\includegraphics[width=1\columnwidth]{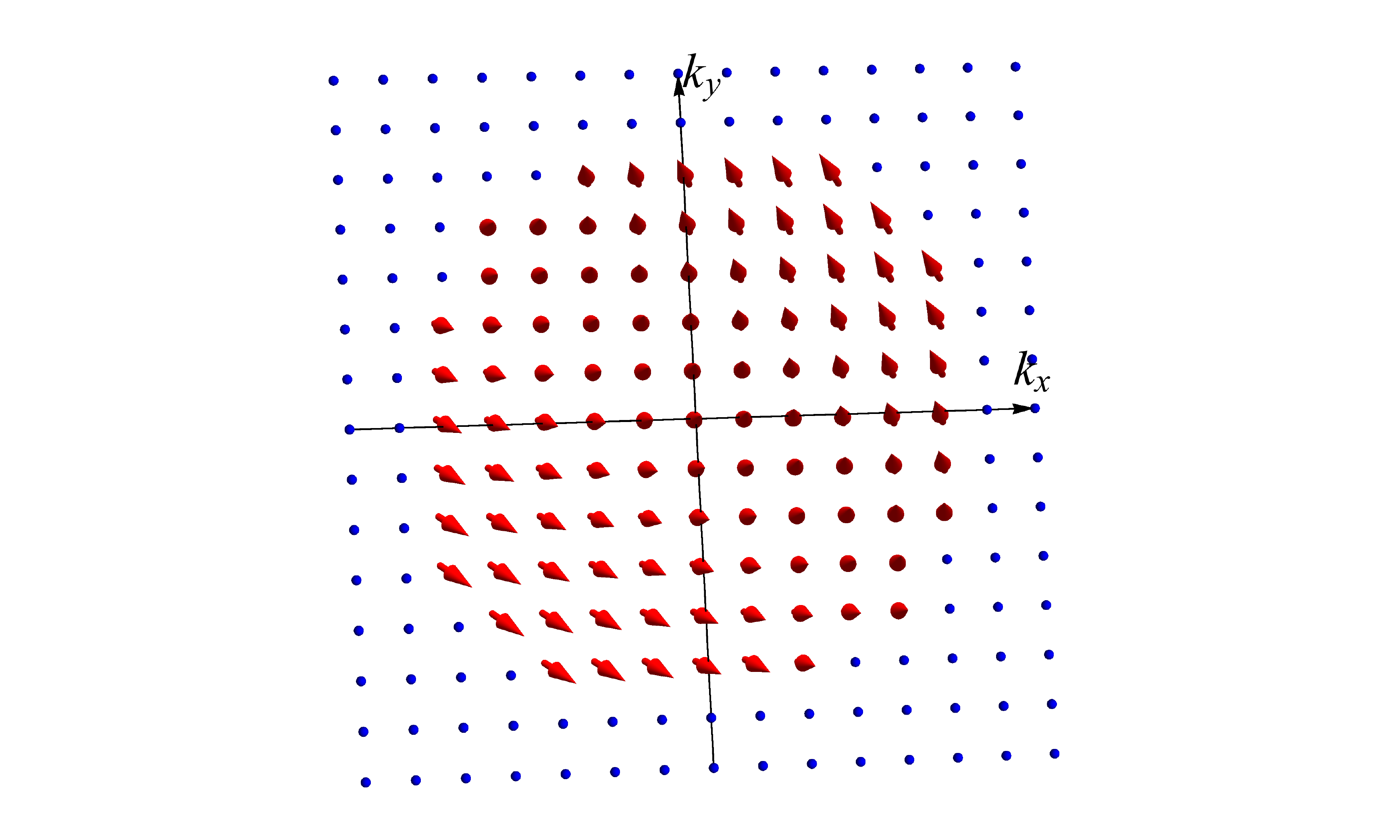}
\caption{\label{OP} OP$^*$ state with $\tilde{\alpha}=0.23$, $\tilde{\beta}=0.23$, $r_s=2.05$.}
\end{center}
\end{figure}

\begin{figure}
\begin{center}
\includegraphics[width=1\columnwidth]{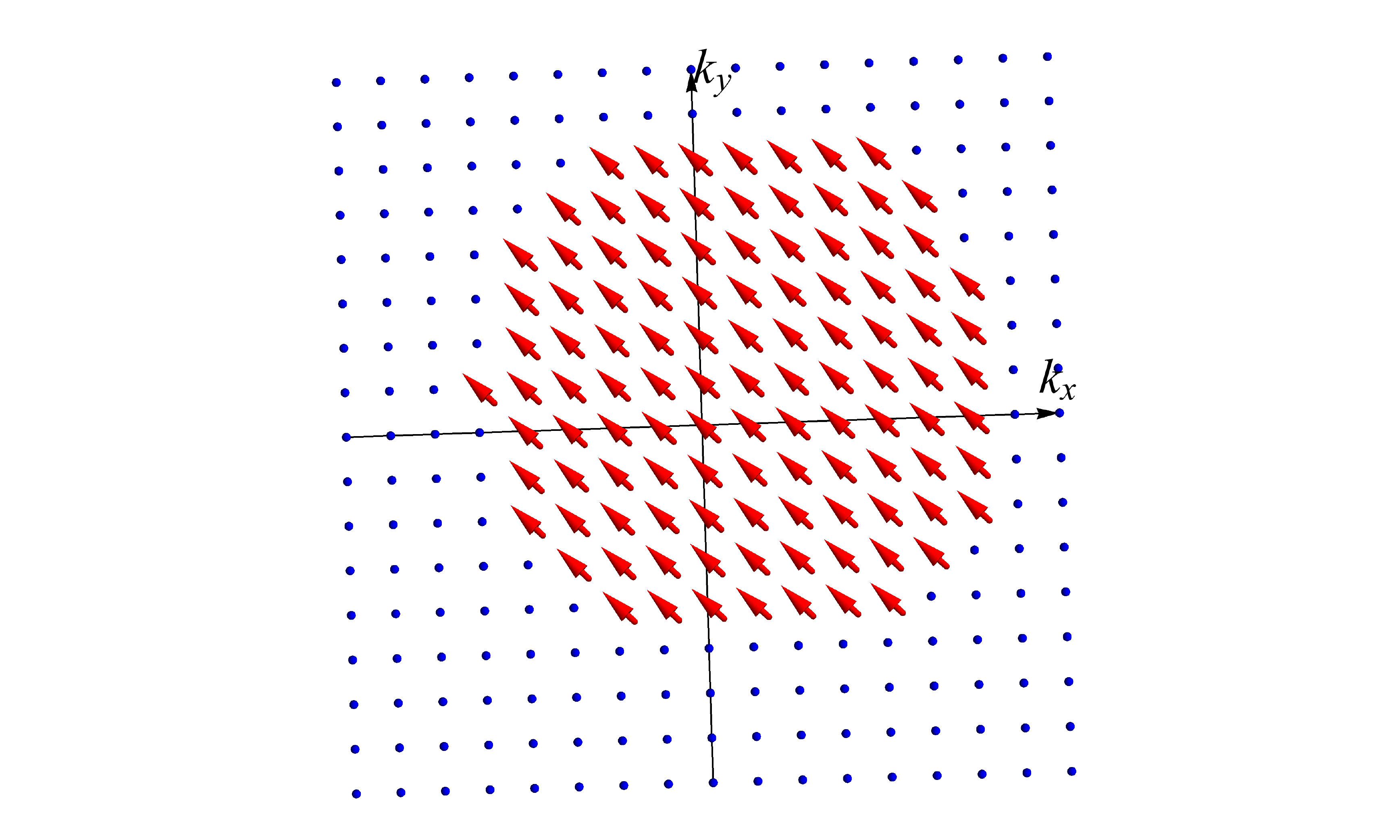}
\caption{\label{IP}IP state with $\tilde{\alpha}=0.23$, $\tilde{\beta}=0.23$, $r_s=2.30$.}
\end{center}
\end{figure}

All realistic systems have both Rashba and Dresselhaus spin-orbit coupling occurring together. In this subsection, we take both coupling with equal strengths into account, where $\tilde{\beta}$ represents the strength of Dresselhaus spin-orbit coupling. Here $\tilde{\beta}$ is a dimensionless variable. We find four phases and plot the phase diagram as functions of $\tilde{\alpha}$ and $r_s$ in Fig.~\ref{B-MCS}. The number of $k$ points is $N=997$. The 2FSs and 2FSs-separate phases are the conventional Fermi liquid (FL) states. The only effect of the Coulomb exchange interaction is to renormalize upwards the strength of the spin-orbit coupling.  The FL state minimizes the single-particle energy by using the non-interacting states and occupation numbers.  When $r_s>2.01$, the MC simulation shows that there is a narrow region which has partially out-of-plane spin-polarization, and we call it the OP$^*$ phase to distinguish it from the OP phase found in Ref.~\cite{Weizhe_Phase_PRB_2017}. The spin texture of the OP$^*$ phase is shown in Fig.~\ref{OP}. For large $r_s$, the right half of the phase diagram is the IP phase shown in Fig.~\ref{IP}. A key feature of the IP phase is that the spin polarization is completely in-plane, as shown in Fig.~\ref{IP}.  The IP state gains exchange energy through the finite polarization and the Fermi surface is shifted. Hartree-Fock simulations are roughly consistent with the analytical results in Sec.~\ref{Ground-states}, but the extremely small energy difference between the IP and the OP$^*$ phases is not so easy to interpret. 

\begin{figure}
\begin{center}
\includegraphics[width=1\columnwidth]{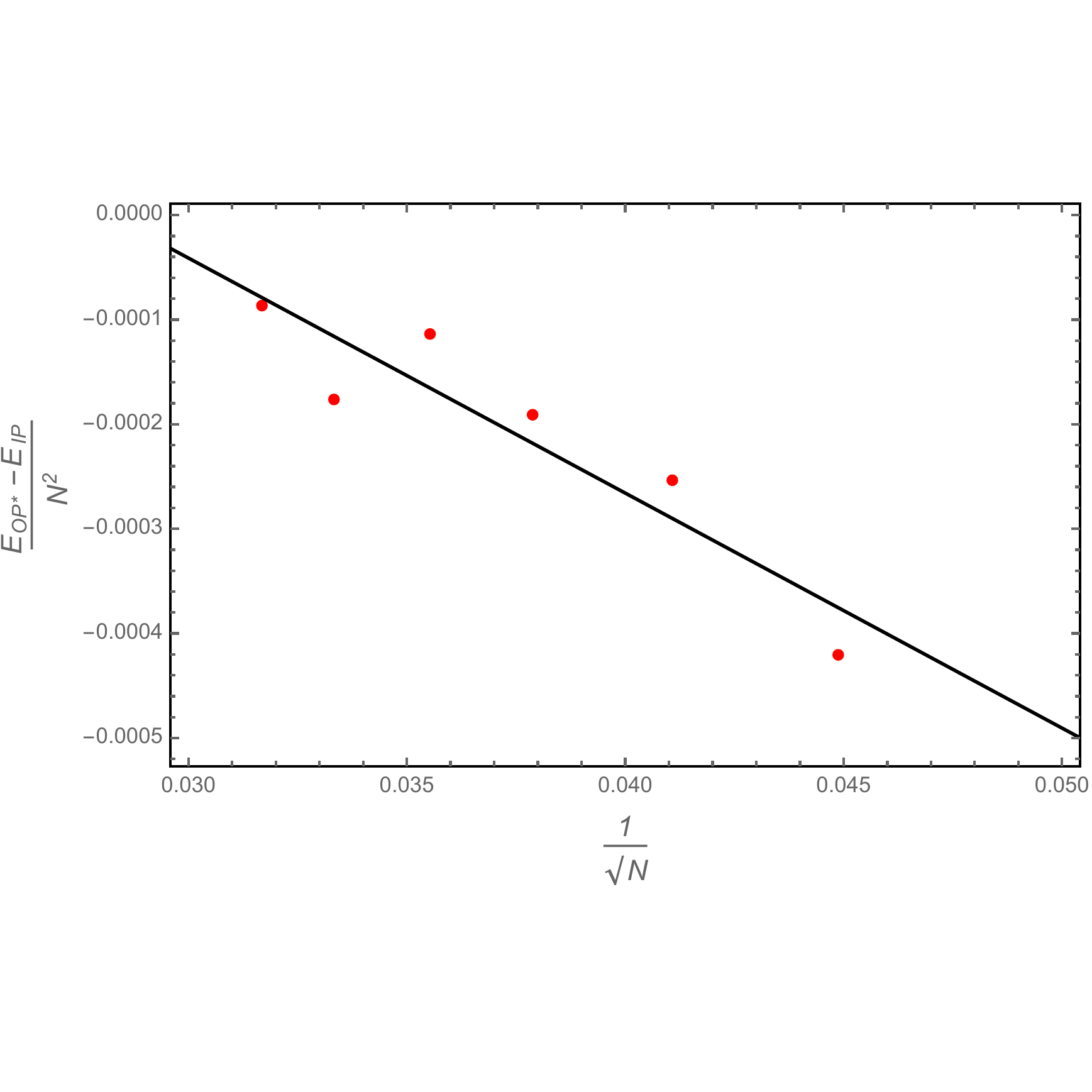}
\caption{\label{Linear-Fit} Linear fit to find the ground state at $\tilde{\alpha}=\tilde{\beta}=0.08$, $r_s=2.10$, where $N$ is the lattice number we take in the Monte-Carlo simulation.}
\end{center}
\end{figure}

To interpret the results from our Monte-Carlo simulation, we choose a small enough spin-orbit parameter $\tilde{\alpha}=0.16$ and different ratios $\tilde{\beta}/\tilde{\alpha}$ to determine the transition points, which are shown in Table~\ref{Shrink-Boundary}. From Table ~\ref{Shrink-Boundary}, we see that the OP$^*$ region is shrinking with increasing ratio 
$\tilde{\beta}/\tilde{\alpha}$ and one possible interpretation is that there is a narrow region where the OP$^*$ phase is the ground state. The classical MC simulation is not sensitive enough to do more than indicate the trend of the phase boundary between the OP$^*$ and IP states, since the discretization error of the ${\bm k}$space is of the same order as the energy difference between the OP$^*$ phase and IP phase. Taking the thermodynamic limit $(N\rightarrow \infty)$, we find that the IP phase has lower energy than the OP$^*$ phase, see Fig.~\ref{Linear-Fit}. So when $r_s>2.01$, the ground state is the IP phase. Our MC code does not capture the spin density wave states, but those states certainly exist \cite{PhysRevB.80.125120,PhysRevB.97.075131}. Based on the analysis from our Monte-Carlo simulation, we believe that the OP$^*$ phase is a remnant of the spin density wave phase. 
\om \begin{table}[H]
	\centering
	\caption{\label{Shrink-Boundary} The boundary with increasing ratio $\tilde{\beta}/\tilde{\alpha}$, where we choose $\tilde{\alpha}=0.16$.}
	\begin{tabular}{|m {2cm}|m{3cm}|m{3cm}|}
		\hline
		$\tilde{\beta}/\tilde{\alpha}$  & 2FSs-OP$^*$& OP$^*$-IP \\
		\hline 
		$0$  & 2.010  & $2.250$ \\
		\hline
		$0.5$ & $1.9959\pm0.0003$  &  $2.2319\pm0.0037$   \\
		\hline
		$1$  & $1.9923\pm0.0001$  &  $2.2072\pm0.0009$\\
		\hline
	\end{tabular}
\end{table}\fi
\begin{table}[H]
	\centering
	\caption{\label{Shrink-Boundary} The boundary with increasing ratio $\tilde{\beta}/\tilde{\alpha}$, where we choose $\tilde{\alpha}=0.16$.}
	\begin{tabular}{|m {2cm}|m{3cm}|m{3cm}|}
		\hline
		$\tilde{\beta}/\tilde{\alpha}$  & 2FSs-OP$^*$& OP$^*$-IP \\
		\hline 
		$0$  & 2.010  & $2.250$ \\
		\hline
		$0.5$ & $1.996$  &  $2.232$   \\
		\hline
		$1$  & $1.992$  &  $2.207$\\
		\hline
	\end{tabular}
\end{table}

\section{Experimental Verification}\label{Experimental Verification}
\subsection{Materials}
The phase transitions presented in this paper occur as a function of electron density $r_s$ and the two parameters $\alpha$ and $\beta$ that characterize the strengths of the Rashba and Dresselhaus interactions.  In 2D systems $r_s$ and $\alpha$ are tunable independently by means of the application of gate voltages and modulation doping.  $\beta$ is more usually thought of as an intrinsic parameter, but even it depends surprisingly strongly on the details of the interface and it may therefore ultimately be variable as well.  Thus the 2D case offers many advantages over the 3D case in the area of tunability.  Indeed, in spite of many years of searching it is still somewhat unclear whether the ferromagnetism long predicted at low density in 3D has been observed, though there are some interesting experimental results along these lines Ref.~[\onlinecite{Young-Nature-1999}].  

In true 2D systems it may be difficult to obtain spin-orbit coupling strengths strong enough for the interesting effects postulated here to occur.  The criterion is that the spin-orbit lengths $\hbar / m \alpha$ or $\hbar / m \beta$ should be comparable to the inter-electron spacing. In Si and SiGe devices the spin-orbit coupling is simply too small.  In GaAs, the Rashba spin-orbit lengths are around $10^{-7}$m while typical devices have interparticle spacings perhaps a factor of 5 less than this.  Working in this material probably the best way to observe the new phases in the near term.  Indeed, there are indications of a spin-polarized state in 2D Si $\delta$-doped GaAs/AlGaAs heterostructures \cite{Ghosh-PRL-2004}.   

 Hole systems are also promising, since the Rashba spin-orbit energy can be as large as $40\%$ of the Fermi energy \cite{Elizabeth-PRB-2017} and second order effects in charge transport can be sizable \cite{PhysRevLett.121.087701}.  The drawback in these systems is disorder.  The mean free path $\ell$ tends to be short and one certainly needs $k_F \ell >> 1$ to observe anything. 

Another intriguing possibility is the gas of surface states on topological insulators.  These are the only systems where a one Fermi surface state has actually been observed to date \cite{Hsieh-Science-2008}.   On the reverse side, disorder seems to be strong also in this case.

\subsection{DC Transport}
In 2D electron systems transport measurements are always the the easiest to carry out.  Since the various transitions that are envisioned here are first order, we expect discontinuous changes in both the longitudinal and the Hall resistances.  The coupled transport equations for charge and spin have been written down for arbitrary $\alpha$ and $\beta$ \cite{Bernevig-ShouCheng-PSH-2006}, but have not yet found a detailed solution.  However, we may understand the qualitative behavior by considering the linearized Boltzmann equation for a simple model of short-range spin-preserving impurity scattering.  We calculate only the longitudinal resistance $\sigma_{ij}$ at zero magnetic field.  To begin with, we focus on the case of $\beta < \alpha $, since in this case all 4 phases are clearly present.  

The conductivity is given by

\begin{equation}\label{Boltz}
	\sigma_{ij}=\frac{e^2}{4 \pi^2} \sum_{ns} \int d^2k \, \tau_{n{\bm k}s} v_{i, n {\bm k}s} v_{j, n {\bm k}s}\, \delta(E_{n {\bm k} s}-E_{F}).
\end{equation}

Here $n$ labels the pieces of the FS, ${\bm k}$ is the wavevector, $ v_{i, n {\bm k} s}$ and 
$\tau_{n {\bm k}s}$ are respectively the i-th component of the velocity and the transport relaxation time of an electron with the indicated quantum numbers.  The delta function pins the integrand to the FS.

The relaxation time is given by
\begin{equation}\label{tau}
\frac{1}{\tau_{n {\bm k} s} }	=\frac{e^2}{4 \pi^2} \sum_{n's'} \int d^2k' \, W_{n {\bm k}s,n'{\bm k}'s'} (1-\cos\theta_{{\bm k},{\bm k}'}).
\end{equation}

For our model the transition rate is
\begin{align}\label{Rate}
W_{n{\bm k}s,n'{\bm k}'s'}& =\frac{2 \pi}{\hbar} \delta(E(n{\bm k}s)-E_{F}) |\langle n{\bm k}s|U|n'{\bm k}'s'\rangle|^2\\
& = \frac{2 \pi n_{imp} u^2}{\hbar}  \delta(E(n{\bm k}s)-E_{F}) |\langle s({\bm k} | s({\bm k}'\rangle|^2
\end{align}

Since the impurity potential $U$ is point-like, the matrix element $u$ is independent of momentum transfer.  The last, and very important, factor is the overlap of the spinors at ${\bm k}$ and ${\bm k}'$.  Note that u has dimensions of energy times length squared in our normalization.

The amplitude for scattering from ${\bm k}$ to ${\bm k}'$ is proportional to the square of the overlap of the spinors at ${\bm k}$ and ${\bm k}'$. In a completely polarized ferromagnetic state this amplitude is unity and the relaxation time $\tau_f$ is independent of ${\bm k}$:

\begin{equation}\label{tauf}
\frac{1}{\tau_{f} }	= \frac{n_{imp} u^2 k_F}{\hbar^2 v_F}. 
\end{equation}
where $k_F$ and $v_F$ are the Fermi wavevector and the Fermi velocity.  We use $\tau_f$ as a benchmark for the relaxation times of the various phases. 

In the FL1 state there is also only one FS, and $\tau$ is again isotropic.  However, the spin texture puts the spin at ${\bm k}$ at a fixed angle from the direction of ${\bm k}$.  We have 

\begin{equation}\label{tauFL1}
\frac{1}{\tau(FL1)}	=\frac{1}{4 \pi^2} \int d^2k \, W_{n {\bm k}s,n'{\bm k}'s'} (1-\cos\theta_{{\bm k},{\bm k}'}) (1+\cos\theta_{{\bm k},{\bm k}'})/2.
\end{equation}

The $ (1+\cos\theta_{{\bm k},{\bm k}'})/2 $ factor is absent in the corresponding expression for $\tau_f$.  Performing the integral we find $1\tau_f = 4/\tau(FL1)$.  The suppression of backscattering in the FL1 state enhances the conductivity dramatically.  The expression for $\tau(FL2)$ is more complicated but the suppression of backscattering is still present.  Hence the FL2 state has a conductivity that is also significantly enhanced over the ferromagnetic state.  In general the conductivities of the FL1 and FL2 states are expected to be rather similar.      

The OP state is rather close to the completely polarized ferromagnetic state in its spin texture, differing in that on the FS the spin angle from the z-axis is $\alpha$.  This yields some amount of backscattering suppression and leads to 
\begin{equation}\label{tauOP}
\frac{1/\tau(OP)}{1 / \tau(FL1)} = \cos^4(\alpha/2) + \sin^4(\alpha/2) - \cos^2(\alpha/2)\, \sin^4(\alpha/2).
\end{equation}
Thus the OP state has a conductivity close to, but slightly greater than, the completely polarized state.

The IP state has a complex texture that is not easily expressed analytically.  However, it is the most interesting in that it breaks rotational symmetry by virtue of the displacement of the FS.  In fact, the conductivity is anisotropic: $\sigma_{xx} \neq \sigma_{yy}$.  For definiteness, say that the FS moves off center along the y direction. At the same time a ferromagnetic moment in the -x direction develops.  If $\alpha$ is small, the state is nearly completely spin polarized and the conductance is low and nearly isotropic. As $\alpha$ increases, the conductivity increases and becomes anisotropic.  In the y-direction the conductivity mainly comes from the states that have ${\bm k \parallel \pm \hat{{\bm y}}}$.  These states have a ferromagnetic character, with the spins pointing mainly along ${- \bm y}$ and backscattering is therefore allowed.  For the states with ${\bm k \parallel \pm \hat{{\bm x}}}$ we have an antiferromagnetic configuration in the following sense: for${\bm k \parallel + \hat{{\bm x}}}$ the spins are in the -y direction, while for ${\bm k \parallel - \hat{{\bm x}}}$ the spins are in the +y direction.  Hence backscattering is suppressed and we find an enhanced conductivity.  Hence overall for the IP phase (1) the conductivity depends strongly on the so coupling, with anisotropy developing as $\alpha$ increases; (2) $\sigma_{xx}>\sigma_{yy}$; (3) the jump in conductivity on passing from the OP to the IP phase is small at low $\alpha$  and increases as $\alpha$ increases; (4) overall, the conductivity is intermediate between the FL and OP states.

These considerations are summarized in Table 2. 
\begin{table}[H]
	\centering
	\caption{\label{Sigma} Conductivity of different phases at small ${\beta}$}
	\begin{tabular}{|m {2.5cm}|m{1cm}|m{1cm}|m{1cm}|m{1.5cm}|}
		\hline
		Phase & FL1 & FL2 & OP & IP \\
		\hline 
		Conductivity  & High & High & Low & Medium \\
		\hline
		Anisotropic? & No & No & No & Yes   \\
		\hline
	\end{tabular}
\end{table}

By means of transport measurements it should therefore be possible not only to detect phase transitions, but also to identify precisely which phases are involved.

In the case of electrons on the surface of topological insulators, similar considerations apply, though it is often difficult to disentangle surface from bulk transport.  However, one may be able to  perform spin-resolved photoemission and observe textures directly, an option that is not usually available in true 2D systems.

\section{Conclusion}\label{Conclusion}
The competition among the kinetic, interaction, and spin-orbit contributions to the electronic energy produces a rich variety of phases in the parameter space of the relative strengths of these energies. When we add the dimension of the relative strength of Rashba and Dresselhaus couplings $\alpha$ and $\beta$, the presence of an additional symmetry when $\alpha = \beta$ adds to the fascination of this physical system.  We treat the symmetric point performing a canonical transformation and add the information so obtained to our MC simulation within the HF approximation.  When $\alpha \neq \beta$ , we identified 4 distinct ground states: 2FSs, 2FSs-Separate, OP* and IP phases, but when the symmetric point is approached, then the OP* phase gets squeezed out. The various phases have different DC transport properties, which aids experimental identification. 

The Coulomb correlation energy increases the effective mass and the absolute value of the correlation energy of the unpolarized 2DEG ground state is greater than its polarized counterpart \cite{PhysRevB.15.2819,PhysRevLett.88.256601}. So even with the correlation energy taken into account, the unpolarized 2FSs and 2FSs-Separate phases still have lower energies than that of the IP phase. With regard to the Pomeranchuk instability, it is an instability in the shape of the Fermi surface of a material with interacting fermions, causing Landau's Fermi liquid theory to break down \cite{Pomeranchuk1,Quintanilla20081279}.  The changes in Fermi surface topology that we observe in the simulations are due to the action of mean fields and this means that that the resemblance to the Pomeranchuk instability is only superficial.

\acknowledgments
This research was supported by the Australian Research Council Centre of Excellence in Future Low-Energy Electronics Technologies (project CE170100039) and funded by the Australian Government. 

\bibliographystyle{apsrev4-1}

%


\end{document}